# Multiferroic CuCrO$_2$ under High Pressure: *In-Situ* X-Ray Diffraction and Raman Spectroscopic Studies

Alka B. Garg*, A. K. Mishra, K. K. Pandey and Surinder M. Sharma

High Pressure and Synchrotron Radiation Physics Division

Bhabha Atomic Research Centre

Mumbai, India 400 085

Abstract

The compression behavior of delafossite compound CuCrO$_2$ has been investigated by *in-situ* x-ray diffraction and Raman spectroscopic measurements upto 23.2 and 34 GPa respectively. X-ray diffraction data shows the stability of ambient rhombohedral structure upto ~ 23 GPa. Material shows large anisotropy in axial compression with *c*-axis compressibility, $\kappa_c$= 1.26×10$^{-3}$(1) GPa$^{-1}$ and *a*-axis compressibility, $\kappa_a$= 8.90×10$^{-3}$(6) GPa$^{-1}$. Our XRD data shows an irreversible broadening of diffraction peaks. Pressure volume data when fitted to 3$^{rd}$ order Birch-Murnaghan equation of state gives the value of bulk modulus, $B_0$ = 156.7(2.8) GPa with its pressure derivative, $B_0'$ as 5.3(0.5). All the observed vibrational modes in Raman measurements show hardening with pressure. Appearance of a new mode at ~24 GPa indicates the structural phase transition in the compound. Our XRD and Raman results indicate that CuCrO$_2$ may be transforming to an ordered rocksalt type structure under compression.

**Keywords:** Delafossite, CuCrO$_2$, High Pressure, X-ray diffraction, Raman Spectroscopy

*Dr. Alka B. Garg

High Pressure & Synchrotron Radiation Physics Division

Bhabha Atomic Research Centre

Mumbai 400 085

Tel: 022-25591308

Email: alkagarg@barc.gov.in

1.  **INTRODUCTION**

In general optically transparent metal oxides are electrical insulators owing to their wide energy band gap and thus good optical and electrical conductivities are mutually exclusive properties. However, transparent conducting oxides (TCO) [1-5] are good electrical conductors and simultaneously transparent to optical photons. Depending upon the donor or acceptor level in the band gap of such a material these are classified as either p-type or n-type TCO. Though the n-type TCOs were reported nearly six decades ago, the discovery of p-type conductivity in $CuAlO_2$ two decades ago has opened up an entirely new area of research called transparent electronics or invisible electronics, where p-n junction made out of two types of TCO could lead to functional windows that transmit visible light yet generate electricity in response to the absorption of photons. Due to this unique combination of optical transparency and electrical conductivity, these materials have found enormous practical applications in devices requiring transparent contacts such as solar cells [6], LCD displays [7], low thermal remittance coatings [8], light emitting diodes [9], electro-chromatic materials for smart windows [10] *etc*. In fact following the discovery of p-type conductivity in $CuAlO_2$, it has been found that there exist a series of $ABO_2$ type compounds (*A*: monovalent cations: Cu, Ag; *B*: trivalent cations: Al, Ga, In, Cr, Fe, Co, lanthanides) which crystallize in delafossite structure and show p-type conductivity ($CuInO_2$ shows bipolar conductivity) [11-13]. The origin of positive carriers in un-doped delafossites is either due to excess oxygen in the interstitials or copper vacancies. Engineering of optical and electronic band structure by doping in copper delafossites makes them useful as photo-catalysts to produce hydrogen by splitting of water [14] and decomposition of toxic waste gases [15]. Magnesium doping in $CuCrO_2$ and $CuScO_2$ leads to an increase in p-type conductivity by a factor of 1000 [16, 17]. *Ab-initio* calculations and selected area electron diffraction (SAED) measurements show improvement in conductivity of chalcogen doped Cu-delafossite [18]. Recently $CuFeO_2$ has been demonstrated to be used as a novel antimicrobial material [19]. $CuCrO_2$ has been discovered as the first Cu- based catalyst for production of chlorine [20].

Delafossite structure, shown in figure 1, consists of edge-connected $BO_6$ octahedra leading to $BO_2$ layers, which are stacked along the *c*-axis of the hexagonal structure. The $BO_2$ layers can be stacked in different ways along the *c* direction, so that delafossites crystallize in the hexagonal 2H (space group: *P6₃/mmc*) or rhombohedral 3R (space group: *R3m*) structures. These $BO_2$ layers are connected together with triangular metallic planes of monovalent element *A*. This $A^+$ cation is linearly two fold-coordinated with oxygen of upper and lower $BO_6$ layers. In the primitive rhombohedral cell there are only four atoms: one *A*, one *B* and two oxygen atoms. However, in the triple hexagonal cell which is conventionally used to describe this structure, *A* and *B* cations occupy, 3a (0,0,0) and 3b (0,0,0.5) Wyckoff positions respectively. The O atoms are situated at 6c (0,0,*u)* positions [21, 22]. Some of these compounds with *B* as magnetic trivalent cation (Fe or Cr) includes $CuFeO_2$, $CuCrO_2$, $AgCrO_2$, belong to the magnetoelectric multiferroics and at low temperature they undergo a series of magnetic phase transitions as a result of geometrical frustration of magnetic ions on triangular lattice [23, 24]. Interestingly a few of these compounds like $CuScO_2$, $CuLaO_2$, $CuInO_2$, $CuAlO_2$ show negative thermal expansion (NTE) behaviour as revealed by neutron scattering measurements [25, 26]. This behaviour is found to be stronger in $CuScO_2$ and $CuLaO_2$ and is attributed to the anharmonicity of linear O-Cu-O bond along the *c*-axis.

Compression is one thermodynamic variable which can tune the various properties of materials due to substantial reduction in their inter-atomic distances. Experimental investigations of a few of the copper based delafossite compounds under high pressure have revealed interesting behavior in terms of their structural, vibrational and electronic properties. Raman scattering measurements on $CuAlO_2$ [27] and $CuGaO_2$ [28] have reported that these compounds transform to unresolved structures above 34 and 26 GPa respectively. Energy dispersive x-ray diffraction and *in-situ* x–ray absorption near edge structure (XANES) measurements on $CuGaO_2$ under pressure confirm changes in copper environment leaving the environment of other cation unchanged [29]. Recent XRD and x-ray absorption spectroscopy (XAS) measurements on single crystals of $CuAlO_2$ reveal the existence of irreversible phase transition beyond 35 GPa [30]. X-ray diffraction, Mössbauer and XAS measurements on $CuFeO_2$ reveals intricate structural/electronic-magnetic pressure induced phase transition [31, 32]. X-ray diffraction and Raman spectroscopic measurements revealed two structural phase transitions at 1.7 and 7 GPa in $LaCuO_2$ under compression with second high pressure phase quenchable on pressure release [33]. According to Takuya et al [34] the compound $CuCrO_2$ is triangular lattice anti-ferromagnet (TLA) which shows spiral spin structure and the ferroelectric polarisation is generated by this spiral spin order showing a strong coupling between ferroelectricity and the spin structure. When Cu ions are replaced with smaller monovalent ions such as Li and Na, the delafossite structure transform into the so called ordered rock-salt structure in which the stacking pattern of triangular lattice plane (TLP) along the c-axis is slightly different from that in the delofossite structure and the compound $LiCrO_2$, does not show spin driven ferroelectricity. The dielectric, ferroelectric and *ac* calorimetric measurements under high pressure on $CuCrO_2$ have revealed that the magnetic transition temperature $T_N$ remarkably increases on pressurization. However, the magnitude of the dielectric anomaly at $T_N$ is suppressed by applying pressure, and the magnitude of the spontaneous polarization below $T_N$ is abruptly suppressed at around 8 GPa [34]. It would be interesting to look for any structural changes with pressure as a possible cause for this suppression.

To understand the experimentally observed compression behaviour of these compounds several computational studies have been carried out. *Ab-initio* band structure calculations and optical absorption measurements on $CuAlO_2$ and $CuScO_2$ thin films indicate the stability of $CuAlO_2$, an indirect band gap semiconductor upto 20 GPa whereas $CuScO_2$ shows phase transition beyond 13 GPa to an unidentified phase [35]. Calculated phonon frequencies for $CuGaO_2$ and $CuAlO_2$ under pressure have indicated the dynamical instability of a transverse acoustic phonon to be the cause of pressure induced phase transition [27, 28] in these materials. Recent first principles calculations determined the critical pressure of transition for delafossite $CuAlO_2$ to a *leaning* delafossite with a higher band gap, to be 60 GPa [36].

Contrary to usual trend observed for a class of materials under pressure, these compounds do not seem to follow any specific trend in terms of phase transition sequence. However, anisotropic axial compressibility seems to be a common feature of all the copper delafossites with *R*3*m* structure resulting into the regularization of oxygen octahedra as compared to the distorted octahedra at ambient conditions [27-34]. To get more insight into the compression behaviour of these compounds, in this manuscript, we report the synthesis, characterization and high pressure investigations of $CuCrO_2$ by *in-situ* x-ray diffraction and Raman spectroscopic measurements. The obtained results will be discussed in context with the available high pressure data on other copper delafossite compounds.

2. **EXPERIMENTAL SECTION**

    2.1 **Sample synthesis**

Polycrystalline sample of $CuCrO_2$ is synthesized by ceramic method with stoichiometric amount of high purity $Cu_2O$ (99.9%) and $Cr_2O_3$ (99.9%). The constituent oxides are heated at 425 K for 24 hrs prior to weighing to remove moisture or any other organic impurity. The reactants were thoroughly ground with an agate mortar and pestle for two hours. Pellets of 12.5 mm diameter and 5 mm thickness were made from this powder by cold pressing. These pellets were heated at 1473 K for 48 hrs in alumina boat in a chamber furnace and air quenched.

## 2.2 Characterizations

### 2.2.1 Powder x-ray diffraction and Raman spectroscopy

As prepared sample is characterized for its single phase formation by powder x-ray diffraction (XRD) using a rotating anode generator (Rigaku-Make) operating at 50 kV and 50 mA current with the Mo $K_\alpha$ ($\lambda$=0.7107 Å) radiation. Highly oriented pyrolitic graphite monochromator with (002) plane is used for selecting the $K_\alpha$ radiation of molybdenum. Structural details were deduced from Rietveld analysis of the diffraction profiles using GSAS software [37].

Raman spectroscopic measurements were carried out using our laboratory based micro Raman set up which has been assembled around a Jobin-Yvon HR-460 single stage spectrograph with a liquid nitrogen cooled spectrum one CCD detector. An edge filter is used to block the Rayleigh line. A diode pumped solid state laser with wavelength 532 nm is used to excite the Raman modes.

## 2.3 High pressure measurements

In-situ high pressure x-ray diffraction measurements were carried out at BL 11 beam line of Indus II synchrotron source at Indore, India [38]. The data was collected in angle dispersive x-ray diffraction (ADXRD) mode, in transmission geometry. The wavelength of x-ray employed and sample to image plate (IP) distance were calibrated using the diffraction pattern of $CeO_2$. For high pressure measurements, a Mao–Bell type diamond anvil cell (DAC) with diamond anvils of culet size 400 μm was used. Tungsten gasket with a central hole of diameter 200 μm pre-indented to a thickness of 60 μm served as high pressure sample chamber. Fine powdered $CuCrO_2$ sample along with copper as internal pressure marker and methanol-ethanol in 4:1 ratio as pressure transmitting medium were loaded in this gasket hole. EOS of copper was used for *in-situ* pressure calibration [39]. Pressure was determined with an accuracy of 0.04 GPa. X-ray powder patterns at various pressures were collected employing x-ray of wavelength 0.7712 Å. Images of the powder diffraction rings were read from the MAR345 image plate detector with the pixel size of 100×100 $\mu m^2$. The images thus obtained were integrated using FIT2D program [40].

For high pressure Raman scattering studies powdered sample was loaded in a 120 μm diameter hole of a pre-indented stainless steel gasket of 80 μm thickness in a modified Mao-Bell type DAC. Small ruby crystals were loaded along with sample for *in-situ* pressure calibration using well known Ruby fluorescence technique [41]. The accuracy of the pressure determination using this technique is 0.03 GPa. Mixture of methanol ethanol in the ratio of 4:1 is used as pressure transmitting medium.

## 3. RESULTS AND DISCUSSIONS

### 3.1 Ambient characterization

One dimensional x-ray diffraction pattern at ambient pressure and temperature along with Rietveld refined data is shown in figure 2. All the observed diffraction peaks could be fitted with rhombohedral structure (space group $R\bar{3}m$) indicating the single phase formation of the compound. Lattice parameters for the ambient phase of $CuCrO_2$ obtained by Rietveld refinement are $a = b = 2.97670(10)$ Å, $c = 17.1113(10)$ Å, $\alpha = \beta = 90°$, $\gamma = 120°$ with unit cell volume $V = 131.306(10)$ Å$^3$, which is in close agreement with the earlier reported value [42]. The Cu and Cr atoms are at fixed positions and the only parameter which is variable is the z parameter for oxygen atom and its refined value is 0.1001(6). R-factors of the refinement are $R_p = 5.37\%$, $R_{wp} = 7.29\%$.

Primitive unit cell of $CuCrO_2$ consists of 4 atoms resulting in 12 normal modes which can be written in terms of irreducible representation as $\Gamma = A_{1g} + E_g + 3A_{2u} + 3E_u$ of which $E_g$ and $A_{1g}$ are Raman active modes. The $E_g$ and $A_{1g}$ modes represent the triangular lattice vibration perpendicular to the *c*-axis and Cu-O bond vibration along the *c*-axis respectively. Figure 3 shows the Raman spectrum of as prepared $CuCrO_2$ recorded at ambient conditions. The spectrum consists of two modes at ~ 453.54(6) cm$^{-1}$ and ~ 702.71(8) cm$^{-1}$ identified as $E_g$ and $A_{1g}$ respectively and they are in good agreement with the earlier reported values [23]. A few weak modes observed around 208 cm$^{-1}$ and 536 cm$^{-1}$ could be non-zone center modes owing to relaxation of Raman selection rules due to *Cu* vacancies or interstitial oxygen similar to those observed in other delafossite compounds such as $CuAlO_2$ [27] and $CuGaO_2$ [28]. Nevertheless one cannot ignore the role of crystal field excitations in the origin of these additional Raman modes as seen in the Raman spectra of some geometrically frustrated compounds [43].

### 3.2 Structural evolution under pressure

Diffraction patterns of $CuCrO_2$ at a few representative pressures, starting from ambient delafossite structure, are shown in figure 4. As marked in the figure, in addition to the diffraction peaks from $CuCrO_2$ sample, peaks from copper and tungsten used as *in-situ* pressure marker and gasket are also observed. At low pressure all the observed diffraction peaks from the sample could be indexed to ambient rhomohedral structure. On increasing the pressure no noticeable changes in the diffraction patterns could be observed except shifting of peaks to higher angles due to lattice compression. This trend is observed till 23.2 GPa (highest pressure reached in the present XRD measurements). Absence of any extra diffraction peak with pressure indicates that the material is structurally stable under compression upto ~ 23 GPa. Refinement of all the diffraction patterns collected at various pressures is carried out to obtain the evolution of lattice parameters under pressure. Figures 5(a)–5(d) show some of the representative fitted XRD pattern at 5.2, 11.6, 23.2 GPa and completely released pressure along with various R-factors respectively. Due to considerable broadening and merging of the diffraction peaks beyond 10 GPa, only profile refinement was carried out which is sufficient to obtain the correct lattice parameter and cell volume. In figure 6 we show the observed changes in normalized lattice parameters *a, c* and *c/a* variation. Error bars in both the axes and c/a ratio have also been plotted along with fitted data. From this data, the isothermal compressibility $\kappa = - 1/l(\partial l/\partial P)_T$ along the *c*- and a- axes are estimated as $\kappa_c$= 1.26×10$^{-3}$(1) GPa$^{-1}$ and $\kappa_a$= 8.90×10$^{-3}$(6) GPa$^{-1}$, respectively. The numbers in the parenthesis represent the estimated error in the fitting. These values indicate a highly anisotropic behaviour of CuCrO$_2$ which is reflected in pressure variation of c/a ratio also. The data upto 13.5 GPa is used for calculating the c-axis compressibility due to considerable scattering in the data beyond this pressure. It is worth mentioning here that almost all copper based delafossite compounds studied under pressure, show similar anisotropic compression behaviour of lattice parameters resulting in more regularization of oxygen octahedra compared to the distorted octahedra at ambient conditions. In this compound intra-layer compressibility is more than the interlayer compressibility causing the variation in intra and inter layer magnetic coupling, and therefore the spin ordering of alternate Cr$^{3+}$ layers may change. This may result in the loss of ferroelectric polarisation possibly in the form of ferroelectric to anti-ferroelctric transition as has been reported by Takuya et al [34]. This could also be a consequence of structural domain rearrangement on application of pressure. A careful analysis of diffraction data reveals that the diffraction peaks from the sample show broadening and to rule out the cause of this broadening as presence of pressure inhomgeniety, we have plotted the full width at half maxima (FWHM) of various peaks of the sample along with copper (used as *in-situ* pressure marker) and tungsten (used as sample chamber) in figure 7. It is clearly observed that the increase in the width of the sample lines is much faster than the Cu or W suggesting that the observed broadening of the sample diffraction peaks is intrinsic or inherent to the sample and not the artefact of measurements (freezing of pressure transmitting medium *etc*.). However, partial contribution to the broadening in the data observed beyond 10 GPa, which is the quasi-hydrostatic limit of the pressure transmitting medium used in the present measurements cannot be ruled out. No signature of decomposition were seen due to large wavelength used to collect the data as has been recently reported by Garg et al in $HoVO_4$ [44]. It is also observed that along with broadening of diffraction peaks from the sample, the intensity of (006) diffraction peak diminishes with respect to the (102) diffraction peak with increasing pressure and beyond 17 GPa the (006) diffraction peak completely disappears. This observation is similar to the one observed for $CuGaO_2$ compound and based on their detailed extended x-ray absorption fine structure (EXAFS) data; this behaviour has been attributed to the presence of preferential orientation effects which is irreversible [29]. The scattering/anomaly observed in the axial compressibility beyond 15 GPa could be either attributed to the effect of pressure transmitting medium or it may be precursor to the observed structural transition described later. The diffraction patterns on pressure release in our experiments, shown in figure 4 (marked by #), indicate the irreversibility of this phenomenon even in $CuCrO_2$ since the intensity of (006) peak does not reappear, similar to the one observed in $CuGaO_2$. The obtained P-V data plotted in figure 8, is fitted to 3$^{rd}$ order Birch- Murnaghan equation of state (B-M EOS) [45], given as

$P = 3/2B_0[(V_0/V)^{7/3} - (V_0/V)^{5/3}][1 - 3/4(4 - B_0') \times \{V_0/V)^{2/3} - 1\}]$,

where, $B_0$ and $V_0$ are the ambient pressure bulk modulus and volume, respectively; V is the volume at pressure P and B' is first derivative of bulk modulus with pressure. The best fit gives $B_0$ to be 156.7(2.8) GPa with $B_0'$ as=5.3(0.5). R-factor of the fitting is 1.49 % along with value of $\chi^2$ as 3.2 %. The maximum deviation in the experimentally determined pressure and fitted EOS pressure is 0.46 GPa. The experimental data below 10 GPa, which is the hydrostatic limit of the pressure transmitting medium used, when fitted to 3$^{rd}$ order BM-EOS gives the value of $B_0$ and $B_0'$ as 153.0(4.1) GPa and 6 respectively. This value of bulk modulus is slightly higher than the recently reported value for $CuCrO_2$ where authors have reported a very large value of $B_0'$ (17.2) [34]. It is to be noted here that such a large value of

$B_0'$ is unrealistic and indicate that the experimental data used to determine the EOS were hindered by experimental problems [46]. To visualize the correctness of the order of BM-EOS used in fitting, the volume–pressure data is transformed into an f–F, i.e. Eulerian strain versus normalized stress plot. [47,48]. For a Birch–Murnaghan EOS, the Eulerian strain is given by $f = 0.5[(V_0/V)^{2/3} -1]$ and the normalized stress is defined as $F = P/[3f(1 + 2f)^{5/2}]$. The f–F plot gives a direct indication of the compression behaviour. If the data points lie on a horizontal line of constant F then $B_0 = 4$ and the data can be fitted with a second-order BM-EOS. If the data lie on an inclined straight line, the data will be adequately described by a third-order BM-EOS. Positive or negative slopes imply $B_0 > 4$ and $B_0 < 4$, respectively. The intercept on the F axis gives the value of ambient pressure bulk modulus, $B_0$. The positive slope in the present data as shown in figure 9 indicates that the pressure derivative of the bulk modulus is larger than 4 which is indeed the case. We would like to further stress that the XRD data presented in ref. 34 is collected using laboratory x-ray source, which may have broader peaks along with large background. The compressibility data obtained on $CuCrO_2$ from present studies is compared with the existing data on other compounds of the family in table 1. Observed c- axis (interlayer) compressibility is comparable with the other compounds in the family however, the a- axis compressibility (intralayer) is slightly larger than the other compounds.

### 3.2 Evolution of vibrational modes under pressure

Figure 10 shows stacked Raman spectra of $CuCrO_2$ at a few representative pressures with letter r indicating the data during pressure release. All the observed Raman modes stiffen with pressure up to 8 GPa. At ~ 8.7 GPa a very weak Raman mode adjacent to $E_g$ mode at ~ 456 cm$^{-1}$ appears. At 13.4 GPa the relative intensity of Raman mode at ~537 cm$^{-1}$ increases drastically while the other Raman mode at ~559 cm$^{-1}$ becomes broad. Up to 21.8 GPa the observed Raman modes stiffen with pressure. At ~ 24.5 GPa the intensity of $E_g$ mode and defect induced Raman modes drop drastically and a new Raman mode (shown by arrow in figure 10) adjacent to $A_{1g}$ mode appears. The new Raman mode red shifts with pressure up to ~34 GPa while the other Raman modes stiffen with a slower rate as shown in figure 11. This implies that there is a phase transition at around 24.5 GPa from rhombohedral to a new high pressure phase. The red shifted new mode corresponding to vibration of Cu-O bond along *c*-axis indicate the lengthening of the Cu-O bond in the new high pressure phase or there may be change in the coordination of copper. This can be understood with the hypothesis that in the new high pressure phase the Cu-O bond (along the c-axis in the parent phase) gets tilted away from the c-axis resulting in an increased bond length causing the stretching mode to be observed at a lower frequency. This is consistent with the observation of changes in copper coordination in $CuGaO_2$ under high pressure [29].

The above discussion indicates that $CuCrO_2$ may be undergoing a transformation from *R*3*m* delafossite type to ordered rock salt type (similar to $LiCrO_2$) structure under compression. Thus the effect of pressure appears to be similar to the substitution of copper by smaller ionic radii atom as observed in case of $Cu(Li)CrO_2$. However, to confirm the structure of high pressure phase XRD experiments need to be extended at still higher pressure. On release of pressure the new high pressure phase continues to exist up to ~ 16 GPa. At 11.6 GPa the Raman modes corresponding to the parent phase reappear and on complete release the observed Raman modes match with that of parent phase. This implies the reversibility of the phase transition observed at ~ 25 GPa. The Gruneisen parameter of the Raman modes, calculated using the bulk modulus of ambient phase, has been shown in table 2. For the parent phase all the Raman modes show positive values of Gruneisen parameters except one at ~ 208 cm$^{-1}$. The highest Gruneisen parameter value for $A_{1g}$ mode is 1.0742.

**Conclusions**

We have investigated the high pressure behavior of technologically important delafossite CuCrO$_2$ by *in-situ* x-ray diffraction and Raman spectroscopic measurements upto 23.2 and 34 GPa respectively in two independent measurements. From the XRD data it is revealed that the compression behavior of *c-* and *a-* axis is highly anisotropic, typical of delafossite family of compounds. The obtained bulk modulus from the pressure *vs.* volume data is 156.7 GPa which is close to the reported values for the similar delafossite compounds. The observed broadening of diffraction peaks, along with diminishing of (006) peaks intensity could be due to preferred orientation effects as have been explained in case of CuGaO$_2$ based on detailed EXAFS data. Raman spectroscopic studies suggest a structural phase transition beyond 24.5 GPa and stiffening of all the observed modes in the ambient structure. In the high pressure phase one of the mode related to Cu-O bond show softening indicating the increase in the bond length or change in copper coordination.

**References**

1.  P. P. Edwards, A. Porch, M. O. Jones, D. V. Morgan, and R. M. Perks, Dalton Trans. 2995 (2004).

2.  A. N. Banerjee, and K. K. Chattopadhyay, Progress in Crystal Growth and Characterization of Materials **50**, 52 (2005).

3.  P. D. C. King, and T. D. Veal, J. Phys.: Condens. Matter **23**, 334214 (2011).

4.  Aron Walsh, Juarez L. F. Da Silva, and Su-Huai Wei, J. Phys.: Condens. Matter **23**, 334210 (2011).

5.  Su Sheng, Guojia Fang, Chun Li, Sheng Xu, and Xingzhong Zhao, Phys. Stat. Sol. a **203**, 1891 (2006).

6.  A. N. Tiwari, G. Khrypunov, F. Kurdzesau, D. L. B¨atzner, A. Romeo, and H. Zogg, Prog. Photovolt., Res. Appl. **12,** 33 (2004).

7.  G. Chae, Japan. J. Appl. Phys. **40**, 1282 (2001).

8.  I. Hamberg, and C. G. Granqvist, J. Appl. Phys. **60**, R123 (1986).

9. A. Porch, D. V. Morgan, R. M. Perks, M. O. Jones, and P. P. Edwards, J. Appl. Phys. **96**, 4211 (2004).

10. C. G. Granqvist, A. Azens, A. Hjelm, L. Kullman, G. A. Niklasson, D. Ronnow, M. Stromme Mattsson, M. Veszele, and G. Vaiva, Solar Energy **63**, 199 (1998).

11. Chesta Ruttanapun, J. Appl. Phys. **114**, 113108 (2013).

12. K. G. Godinho, B. J. Morgan, J. P. Allen, D. O. Scanlon, and G. W. Watson, J. Phys.: Condens. Matter **23**, 334201 (2011).

13. William C. Sheets, Emmanuelle Mugnier, Antoine Barnabé, Tobin J. Marks, and Kenneth R. Poeppelmeier, Chem. Mater. **18**, 7 (2006).

14. M. N. Huda, Y. Yan, and M. M. Al-Jassim, J. Appl. Phys. **109,** 113710 (2011).

15. K. Gurunathan, J. O. Baeg, S. M. Lee, E. Subramanian, S. J. Moon, and K. J. Kong Catal. Commun. **9**, 395 (2008).

16. R. Nagarajan, A. D. Draeseke, A. W. Sleight, and J. Tate, J. App. Phys. **89**, 8022 (2001).

17. R. Kykyneshi, B. C. Nielsen, J. Tate, J. Li, and A. W. Sleight, J. Appl. Phys. **96**, 6188 (2004).

18. Nilesh Mazumder, Dipayan Sen, Uttam K. Ghorai, Rajarshi Roy, Subhajit Saha, Nirmalya S. Das, and Kalyan K. Chattopadhyay, J. Phys. Chem. Lett. **4,** 3539 (2013).

19. Xiaoqing Qiu, Min Liu, Kayano Sunada, Masahiro Miyauchi, and Kazuhito Hashimoto, Chem. Commun. **48**, 7365 (2012).

20. Amol P. Amrute, Gaston O. Larrazabal, Cecilia Mondelli, and Javier Perez-Ramirez, Angew. Chem. Int. Ed. **52**, 9772 (2013).

21. T. Arima, J. Phys. Soc. Jpn. **76,** 073702 (2007).

22. G. Ehlers, A. A. Podlesnyak, M. Frontzek, R. S. Freitas, L. Ghivelder, J. S. Gardner, S. V. Shiryaev and S. Barilo, J. Phys.: Condens. Matter **25,** 496009 (2013).

23. O. Aktas, K. D. Truong, T. Otani, G. Balakrishnan, M. J. Clouter, T. Kimura, and G. Quirion, J. Phys.: Condens. Matter **24,** 036003 (2012).

24. F. Ye, Y. Ren, Q. Huang, J. A. Fernandez-Baca, Pengcheng Dai, J. W. Lynn, and T. Kimura, Phys. Rev. B **73,** 220404 (2006).

25. J. Li, A. W. Sleight, C. Y. Jones, and B. H. Toby, J. Solid State Chem. **178**, 285 (2005).

26. J. Li, A. Yokochi, T. G. Amos, and A. W. Sleight, Chem. Mater. **14**, 2602 (2002).

27. J. Pellicer-Porres, D. Martínez-García, A. Segura, P. Rodríguez-Hernández, A. Muñoz, J. C. Chervin, N. Garro, and D. Kim, Phys. Rev. B **74**, 184301 (2006).

28. J. Pellicer-Porres, A. Segura, E. Martínez, A. M. Saitta, A. Polian, J. C. Chervin, and B. Canny, Phys. Rev. B **72**, 064301 (2005).

29. J. Pellicer-Porres, A. Segura, Ch. Ferrer-Roca, D. Martı́nez-Garcı́a, J. A. Sans,,E. Martı́nez, J. P. Itié, A. Polian, F. Baudelet, A. Muñoz, P. Rodrı́guez-Hernández, and P. Munsch, Phys. Rev. B **69**, 024109 (2004).

30. J. Pellicer-Porres, A. Segura, Ch. Ferrer-Roca, A. Polian, P. Munsch, and D. Kim, J. Phys.: Condens. Matter **25**, 115406 (2013).

31. W. M. Xu, G. Kh. Rozenberg, M. P. Pasternak, M. Kertzer, A. Kurnosov, L. S. Dubrovinsky, S. Pascarelli, M. Munoz, M. Vaccari, M. Hanfland, and R. Jeanloz, Phys. Rev. B **81,** 104110 (2010).

32. T. R. Zhao, M. Hasegawa, H. Takei, T. Kondo, and T. Yagi, Jpn. J. Appl. Phys. **35**, 3535 (1996).

33. Nilesh P. Salke, Alka B. Garg, Rekha Rao, S. N. Achary, M. K. Gupta, R. Mittal, and A. K. Tyagi, J. Appl. Phys. **115**, 133507 (2014).

34. Takuya Aoyama, Atsushi Miyake, Tomoko Kagayama, Katsuya Shimizu, and Suyoshi Kimura, Phys. Rev. B **87**, 094401 (2013).

35. S. Gilliland, J. Pellicer-Porres, A. Segura, A. Muñoz, P. Rodríguez-Hernández, D. Kim, M. S. Lee, and T. Y. Kim, Phys. Status Solidi B **244**, 309 (2007).

36. A. Nakanishi, and H. Katayama-Yoshida, J. Phys. Soc. Jpn. **80**, 024706 (2011).

37. A. C. Larson, and R. B. Von Dreele 2000 GSAS: General Structure Analysis System Los Alamos National Laboratory, Report LAUR 86-748.

38. K. K. Pandey, H. K. Poswal, A. K. Mishra, Abhilash Dwivedi, R. Vasanthi, Nandini Garg, and Surinder M. Sharma, Pramana- Journal of physic **80**, 607 (2013).

39. W.T. Carter, S.P. Marsh, J. N. Fritz and R. G. McQueen, in Accurate Characterization of the High Pressure Environment, ed. E. C. Lloyd, NBS special pub. 326, Washington DC, p. 147 (1971).

40. A. P. Hammersley, S. O. Svensson, M. Hanfland, A. N. Fitch, and D. Hausermann, High Press. Res. **14**, 235 (1996).

41. G. J. Piermarini, S. Block, and J. D. Barnett, J. Appl. Phys. **44**, 5377 (1973).

42. O. Crottaz, F. Kubel, Zeitschrift fuer Kristallographie **211**, 482 (1996).

43. T. T. A. Lumen, Phys. Rev. B **77**, 214310 (2008).

44. Alka B. Garg, D. Errandonea, P. Rodríguez-Hernández, S. López-Moreno, A. Muñoz, and C. Popescu, J. Phys.: Cond. Matter **24**, 265402 (2014).

45. F. Birch, J. Geophys. Res. **83**, 1257 (1978).

46. D. Errandonea, A. Munoz, and J. Gonzalez-Platas, J. Appl. Phys. **115**, 216101 (2014).

47. Dion L. Heinz, and Raymond Jeanloz, J. Appl. Phys. **55**, 885 (1984).

48. D. Errandonea, R. S. Kuamr, O. Gomis, F. J. Manjon, V. V. Ursaki, and I. M. Tiginyanu, J. Appl. Phys. **114**, 233507 (2013).

Figure captions

Figure 1  Delafossite structure. Linear bonding between Cu and O atom is clearly seen.

Figure 2  Rietveld refined ambient pressure and temperature XRD pattern of as synthesized $CuCrO_2$ showing single phase formation of the compound. Fitted background along with difference plot is also plotted. Vertical tick marks represent allowed reflection of delafossite structure with $R$3m space group.

Figure 3  Raman spectrum of as synthesized $CuCrO_2$ at ambient pressure and temperature conditions.

Figure 4  Evolution of XRD data at several pressures for $CuCrO_2$ along with a few pressure points while pressure unloading. * indicate the diffraction peak from high pressure sample chamber (W). Diffraction peaks from *in-situ* pressure calibrant (Cu) are also indicated. Numbers on the right hand side of y-axis denotes the pressure in GPa.

Figure 5  Observed, calculated and difference plot of x-ray powder patterns for $CuCrO_2$ (a) at 5.1 GPa, (b) 11.6 GPa, (c) 23.2 GPa and (d) pressure released. Top, middle and bottom vertical marks indicate Bragg reflections from the sample, pressure calibrant (Cu) and sample chamber (W) respectively. Fitted background along with difference plot is also plotted.

Figure 6  Pressure dependence of (a) normalized cell parameters and (b) axial ratio of $CuCrO_2$. Symbol presents the experimental data points and solid line is BM-EOS fit to data. Error bars have also been plotted in the data.

Figure 7  Full width at half maxima (FWHM) of various diffraction peaks from the sample along with prominent peaks from pressure chamber (W) and pressure calibrant (Cu).

Figure 8    Volume data for $CuCrO_2$. Solid circles are experimental data points and solid line is the $3^{rd}$ order Birch–Murnaghan equation of state fit to all data points. Errors in the volume have also been plotted and they are smaller than the symbol used. Dotted line shows the $3^{rd}$ order BM-EOS fit to the experimental data below 10 GPa.

Figure 9    Volume–pressure data of $CuCrO_2$ displayed as a plot of the normalized pressure F against the Eulerian strain f. Dependence of F and f on pressure and volume data has been described in the text. Solid squares are the experimental points whereas solid line is linear fit to the data.

Figure 10   Stacked Raman spectra of $CuCrO_2$ at a few representative pressures.

Figure 11   Variation of Raman frequencies shifts of $CuCrO_2$ with pressure. Error bars are smaller than the symbol size.

Table 1

Experimental compressibility data of copper based delafossite family of compounds

| Compound | | $B_0$ (GPa) | $B_0'$ | $\kappa_a (10^{-3} \text{ GPa}^{-1})$ | $\kappa_c (10^{-3} \text{ GPa}^{-1})$ |
|---|---|---|---|---|---|
| $CuCrO_2$ | [34] | 126.8 | 17.1 | 2.30(57) | 0.392(866) |
| | This work | 156.7(2.8) | 5.3(0.5) | 8.90(6) | 1.26(1) |
| $CuAlO_2$ | [30] | 200(10) | - | 2.06(5) | 0.83(4) |
| $CuGaO_2$ | [29] | 202(15) | - | 1.96(5) | 0.75(4) |
| $CuLaO_2$ | [33] | 154(25) | 4.8 | 2.5(1) | 1.04(7) |
| $CuFeO_2$ | [31] | 148.0(0.7) | 4 | - | - |
| | [32] | 156 | 2.6 | 2.58(4) | 0.65(2) |

Table 2

Experimentally observed Raman modes along with Gruneisen parameters (γ)

Numbers in the parenthesis denotes the error in the respective parameter.

| Raman Mode ($\omega$) (cm$^{-1}$) | Assignment | dw/dp (cm$^{-1}$/GPa) | Gruneisen Parameter ($\gamma$) |
|---|---|---|---|
| 207.73(15) | $E_g$ | -0.32(9) | -0.24(1) |
| 453.54(6) | $E_g$ | 2.55(6) | 0.88(4) |
| 535.76(120) | Defect Induced | 2.89(7) | 0.85(5) |
| 585.62(290) | Defect Induced | 3.29(16) | 0.88(12) |
| 702.71(8) | $A_{1g}$ | 4.79(11) | 1.07(15) |

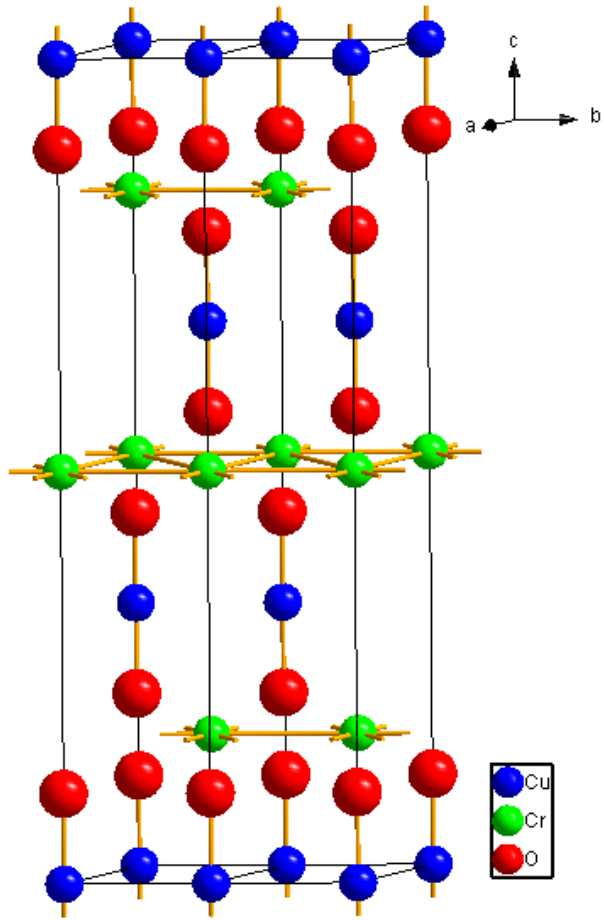

Figure 1 Garg et al

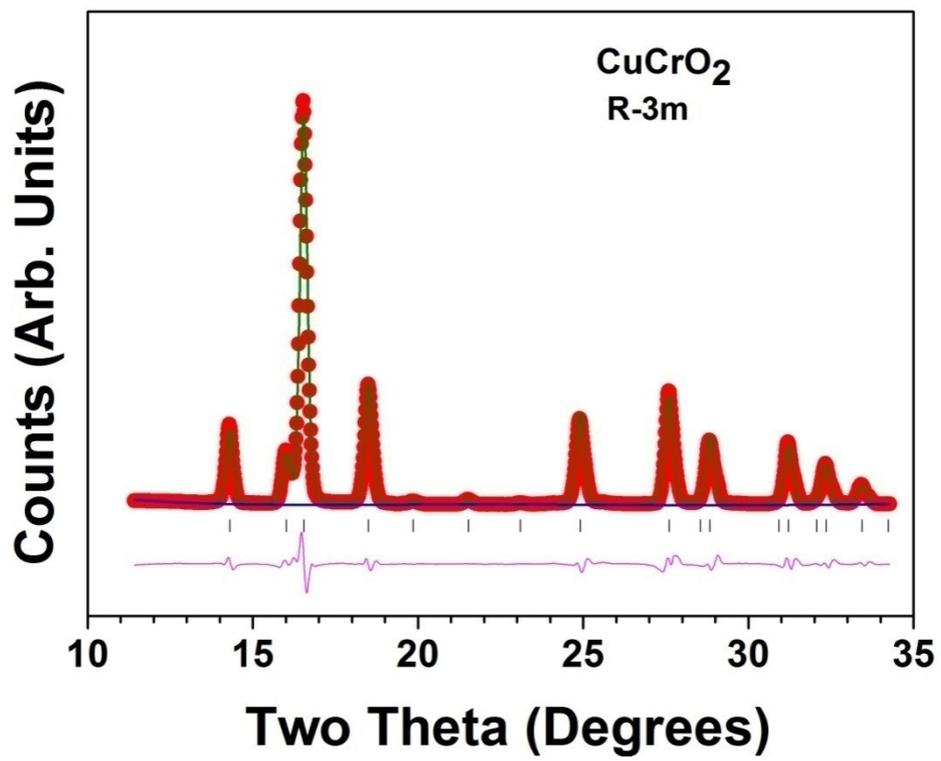

Figure 2 Garg et al

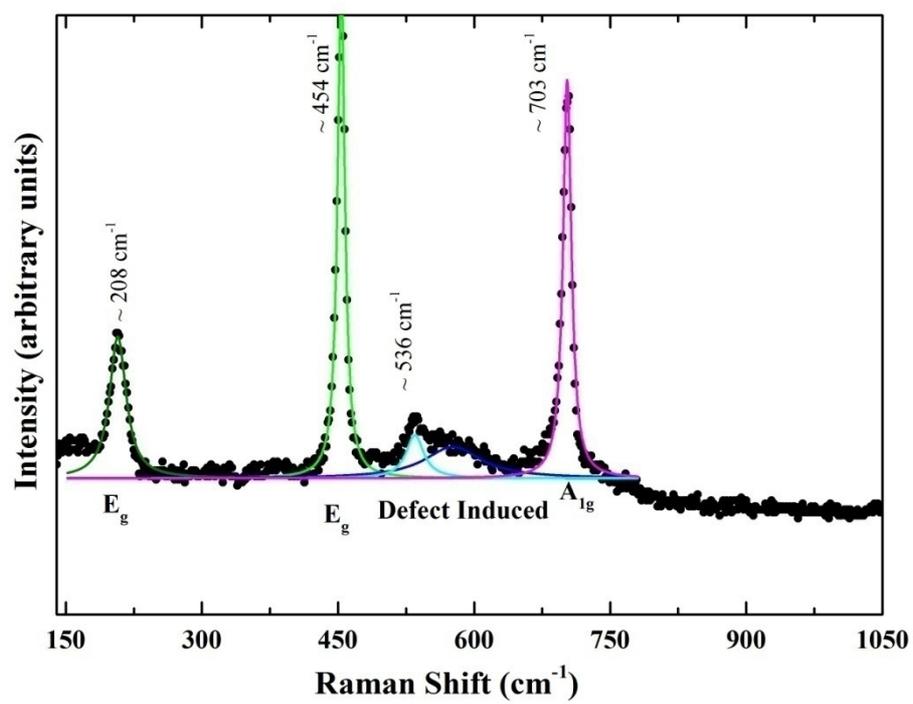

Figure 3: Garg et al

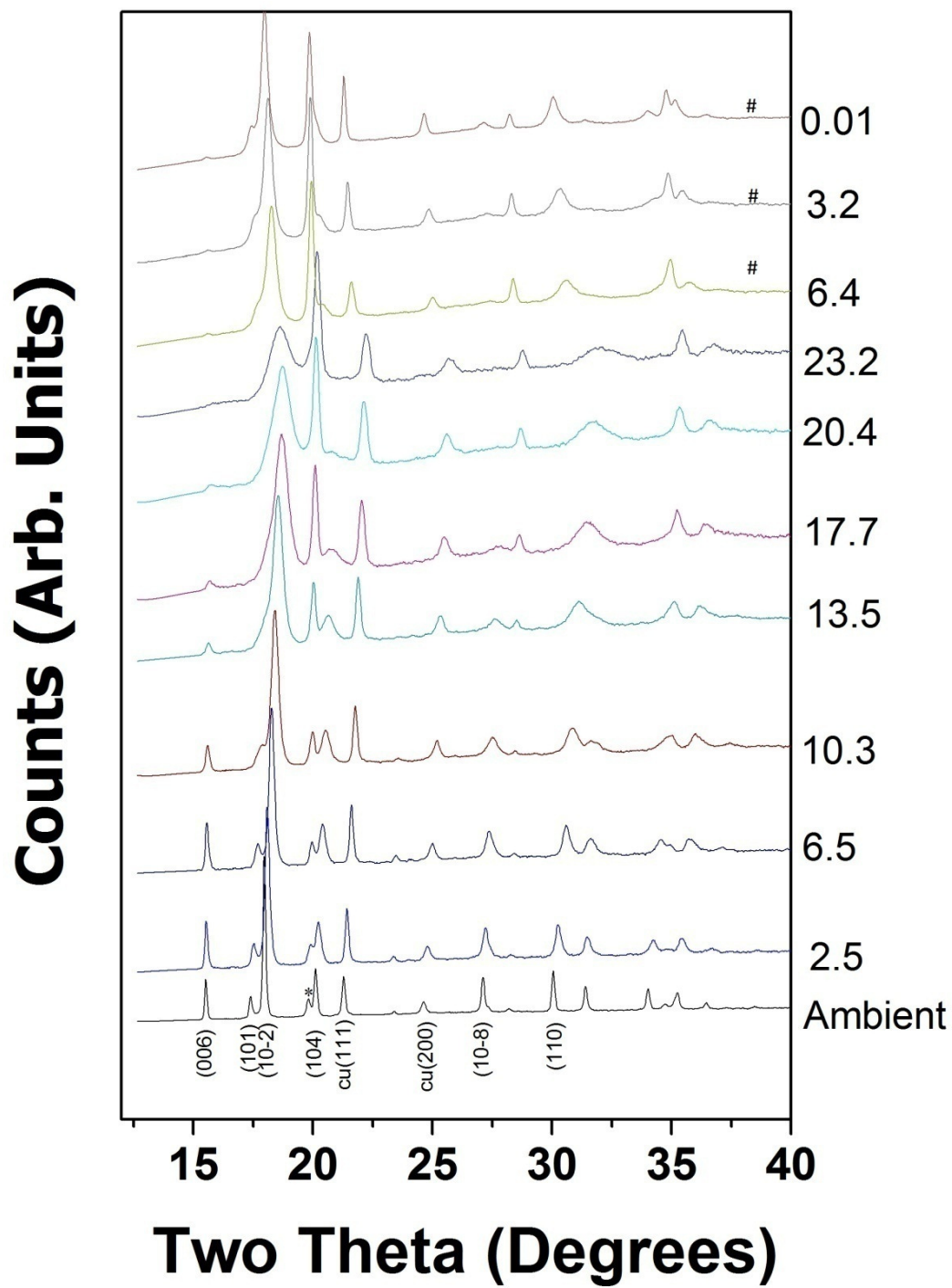

Figure 4 Garg et al

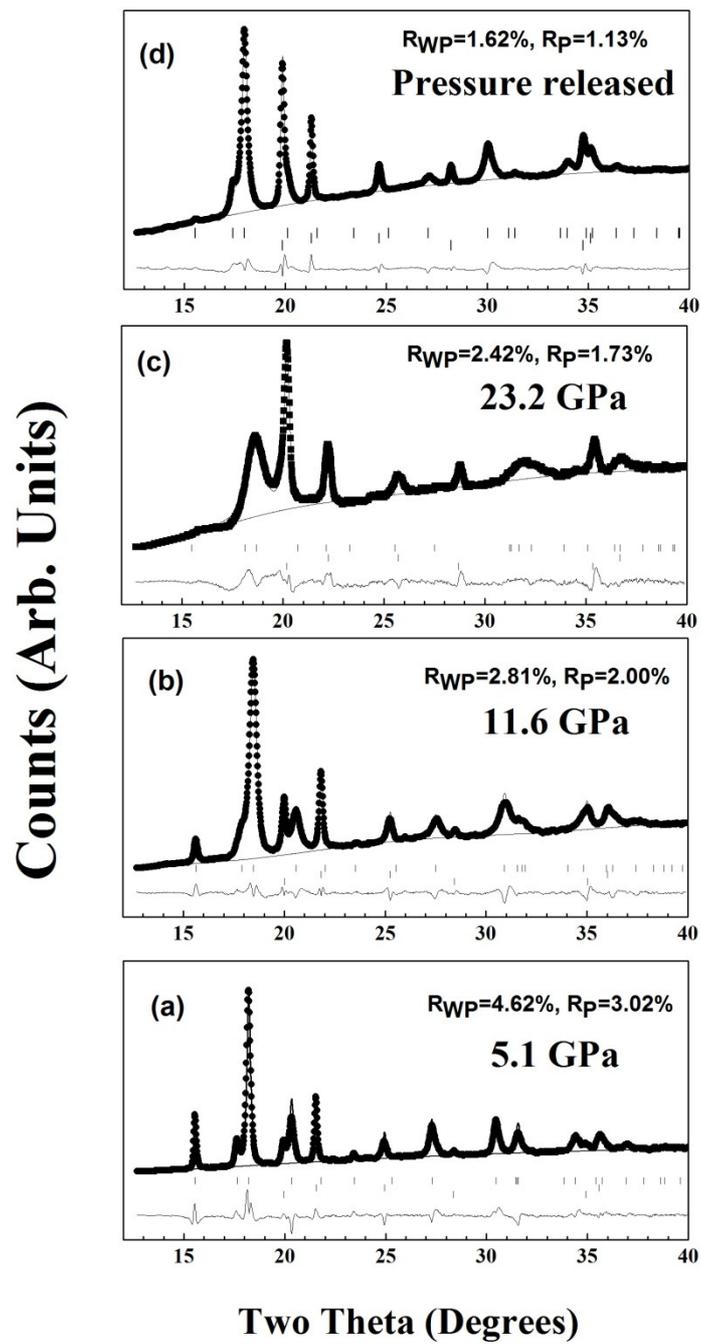

Figure 5 Garg et al

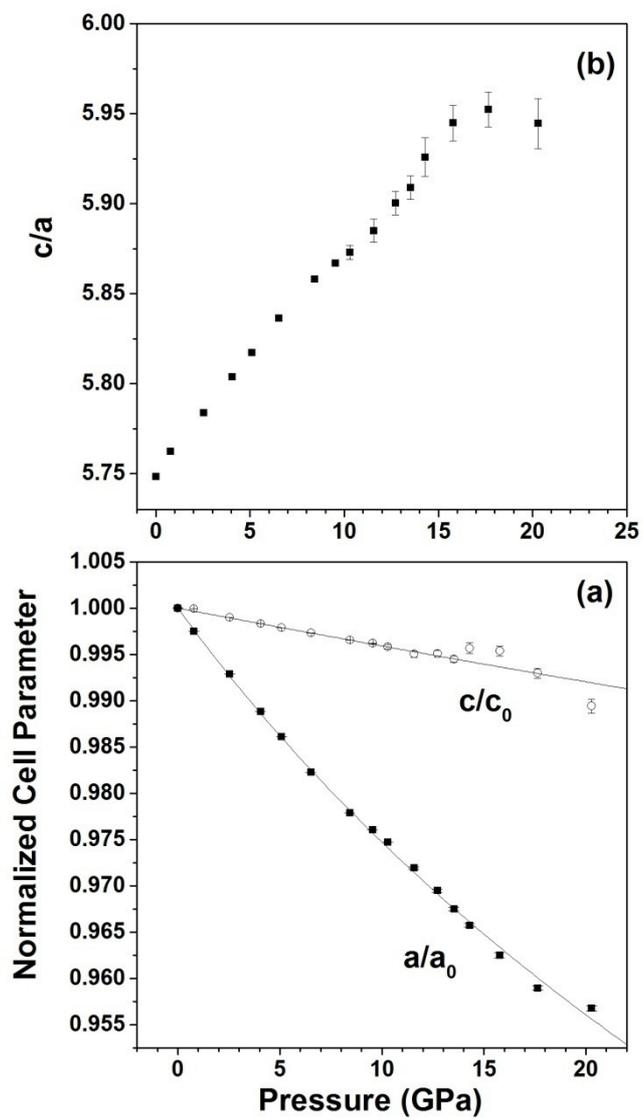

Figure 6 Garg et al

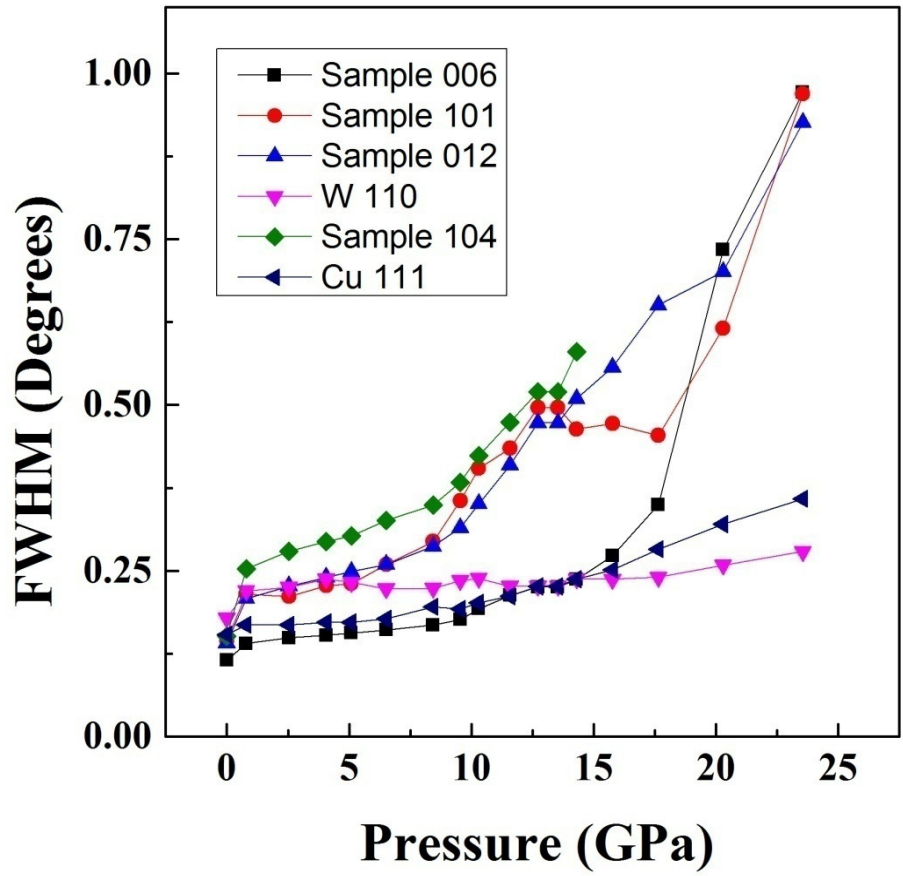

Figure 7  Garg et al

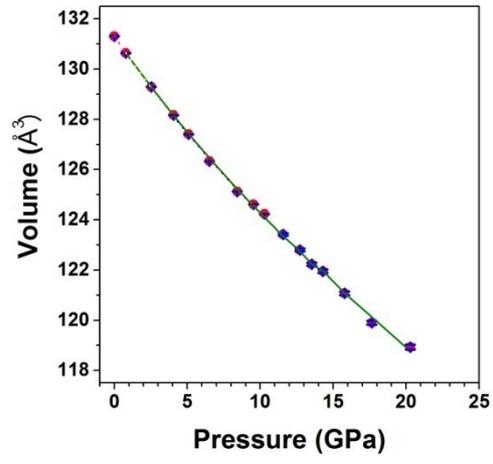

Figure 8 Garg et al

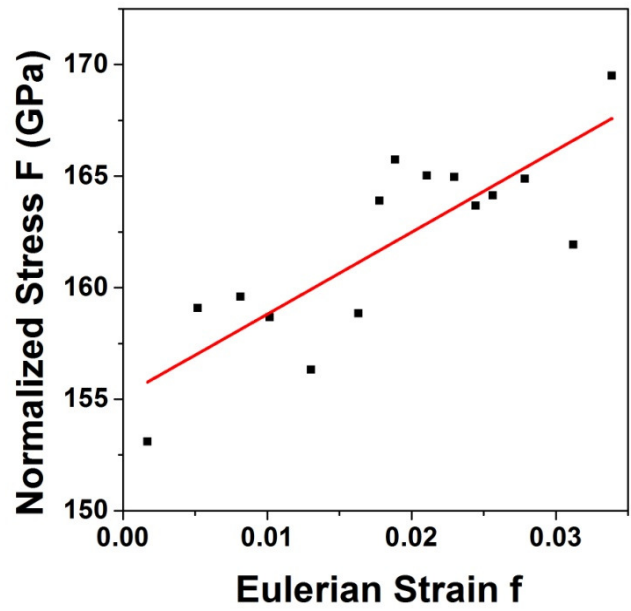

Figure 9  Garg et al

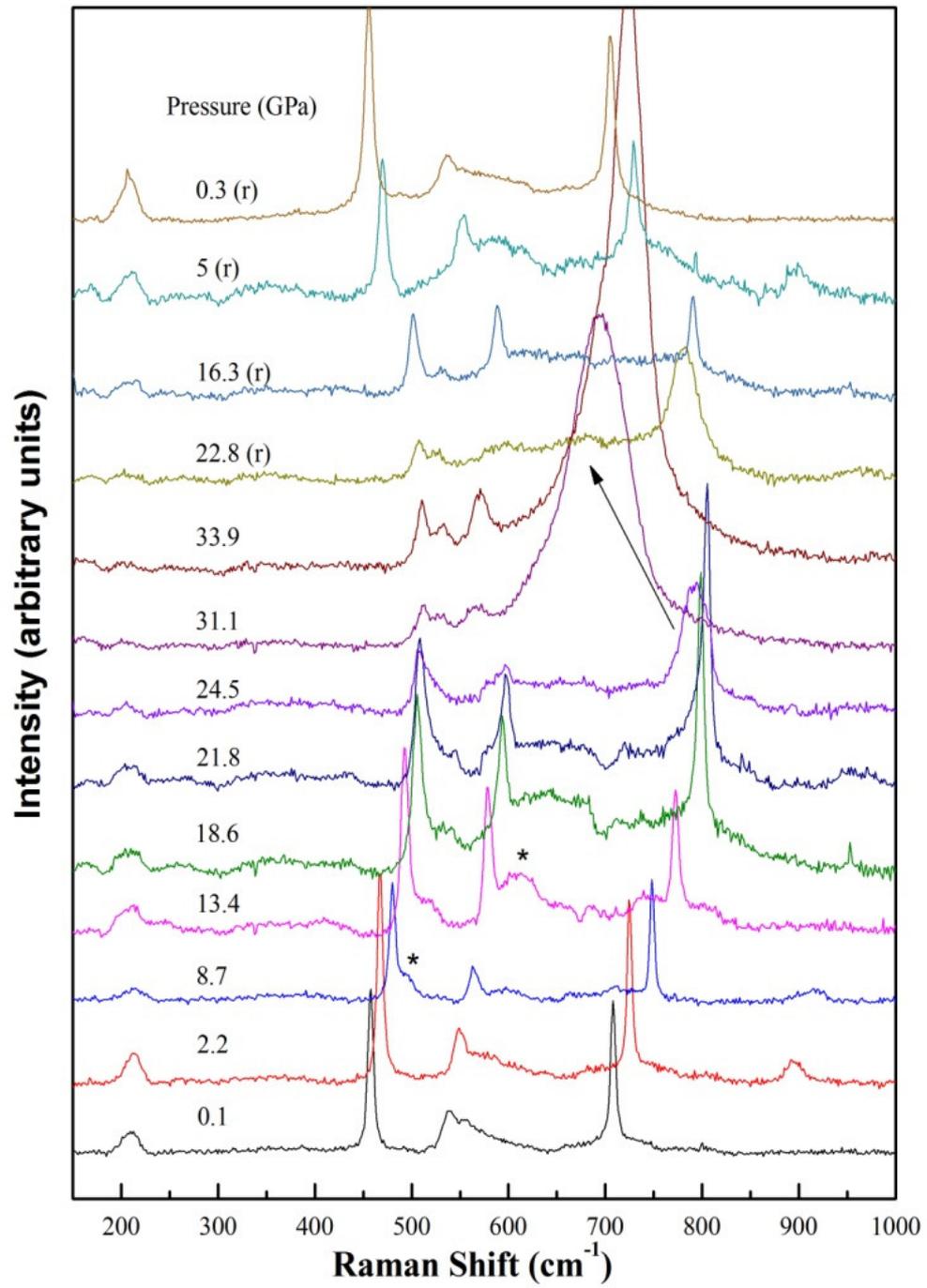

Figure 10 Garg et al

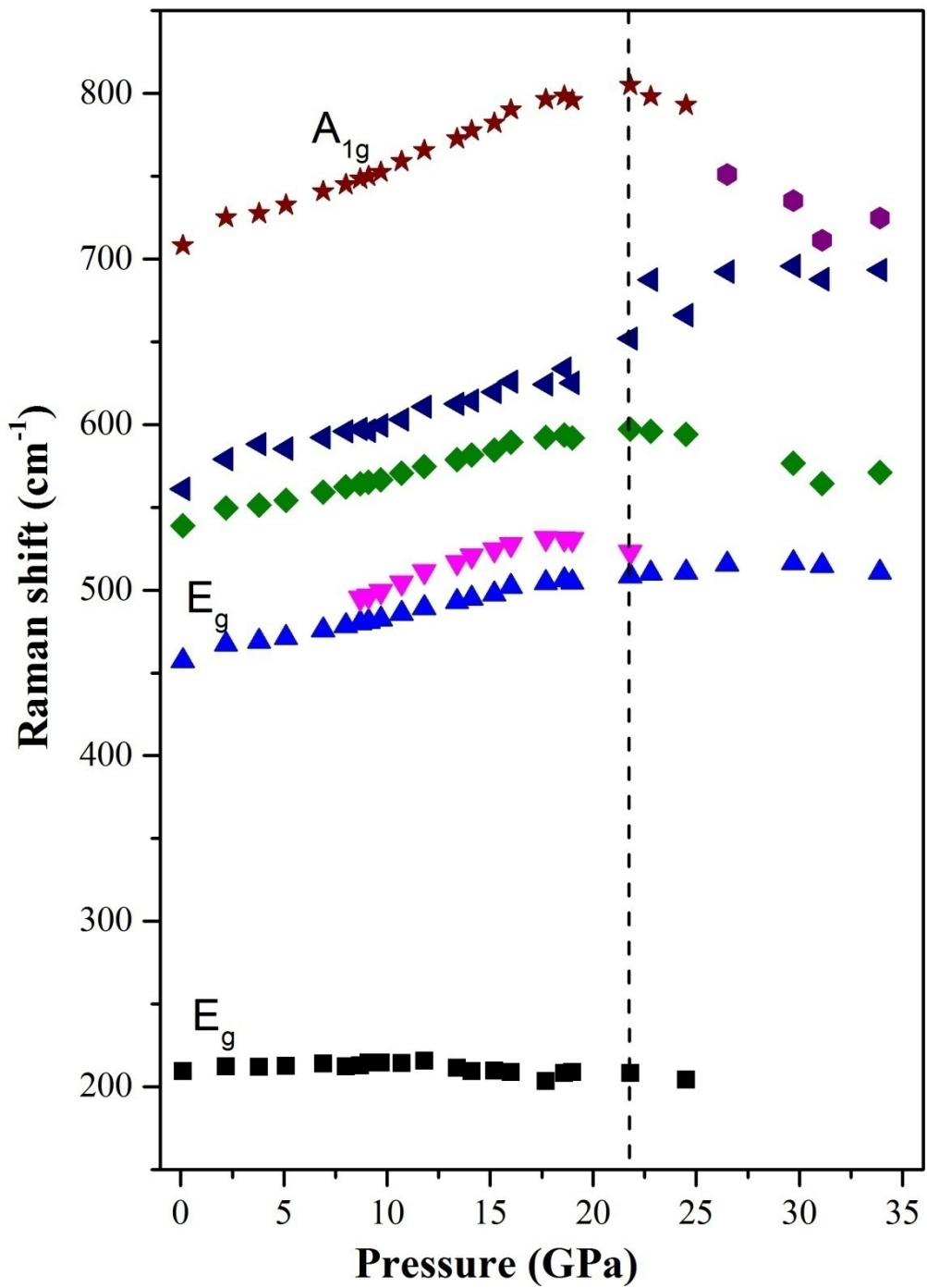

Figure 11  Garg et al